\documentclass[superscriptaddress,twocolumn,pra,showpacs]{revtex4}

\usepackage{amsmath,amssymb}
\usepackage{graphics}
\usepackage{epsfig}

\begin{document}

\title{ Origin of the Three-body Parameter Universality in Efimov
  Physics}

\author{Jia Wang}
\affiliation{Department of Physics and JILA, University of Colorado,
Boulder, CO 80309-0440}

\author{J. P. D'Incao}
\affiliation{Department of Physics and JILA, University of Colorado,
Boulder, CO 80309-0440}

\author{B. D. Esry}
\affiliation{Department of Physics, Kansas State University,
Manhattan, Kansas 66506, USA} 

\author{Chris H. Greene}
\affiliation{Department of Physics and JILA, University of Colorado,
Boulder, CO 80309-0440}

\begin{abstract}
In recent years extensive theoretical and experimental
  studies of universal few-body physics have led to advances in our
  understanding of universal Efimov physics. Whereas theory 
  had been the driving force behind our understanding of
  Efimov physics for decades, recent experiments have contributed an
  unexpected discovery. Specifically, measurements have
  found that the so-called {\em three-body parameter} determining several  
  properties of the system is universal, even though fundamental
  assumptions in the theory of the Efimov effect suggest that it
  should be a variable  property that depends on the precise  details of
  the short-range two- and three-body  interactions. The present
  Letter resolves this apparent contradiction by elucidating 
  previously unanticipated 
  implications of the two-body interactions. Our study shows that
  the three-body parameter universality emerges because a universal
  effective barrier in the  three-body potentials prevents the three
  particles from simultaneously getting close together. Our
  results also show limitations on this universality, as it is more
  likely to occur for neutral atoms and less  likely to extend to
  light nuclei.
\end{abstract}

\maketitle 

In the early 70's, Vitaly Efimov predicted a
strikingly counterintuitive quantum phenomenon \cite{Efimov}, today
known as Efimov effect: in three-body systems for which the two-body
$s$-wave scattering length $a$ is much larger than the characteristic
range $r_{0}$ of the two-body interaction, an infinite number
of three-body bound states can be formed even when the short-range
two-body interactions are too weak to bind a two-body state ($a<0$). 
The Efimov effect, once considered a mysterious and
  esoteric effect, is today a reality that many experiments in
  ultracold quantum gases have successfully observed and  continued to
  explore \cite{133Cs_IBK_1,133Cs_IBK_2,39K_LENS,7Li_Rice,7Li_Kayk_A,7Li_Kayk_B,6Li_Selim_A,6Li_Selim_B,
 6Li_PenState_A,6Li_PenState_B,85Rb_JILA,RbK_LENS,UedaExp}. 
One of the most fundamental assumptions underlying our theoretical
understanding of this peculiar effect is that the weakly bound
three-body energy spectrum, and  other low-energy three-body scattering
observables, should depend on a three-body parameter that encapsulates
all details of the interactions at short distances \cite{BraatenReview}.
For this reason, the three-body parameter has been viewed as 
nonuniversal since its value for any specific system would depend on the
precise details of the underlying two- and three-body
interactions \cite{3BodyForces,3BodyForcesN,DIncaoJPB}. 

In nuclear physics, this picture seems be consistent, i.e., three-body
weakly bound state properties seem to be sensitive to the nature of the
two- and three-body short-range interactions \cite{3BodyForcesN}. More
recently, however, Berninger {\em et al.} \cite{133Cs_IBK_2} have
directly explored this issue for alkali atoms whose scattering lengths
are magnetically tuned near different Fano-Feshbach resonances
\cite{ChinReview}. 
Even though the short-range physics can be expected to vary from one resonance to another, 
Efimov resonances were found for values of the magnetic 
field at which $a$=$a^{-}_{\rm 3b}$=$-9.1(2) r_{\rm vdW}$, 
where $r_{\rm vdW}$ is the van der Waals length \cite{CommentAmean,Flambaum}. 
Therefore, in each of these cases, the three-body parameter was approximately the
same, thus challenging a fundamental assumption of the universal theory. 
Even more striking is the observation that the
Efimov resonance positions obtained for $^{39}$K \cite{39K_LENS},
$^{7}$Li~\cite{7Li_Rice,7Li_Kayk_A,7Li_Kayk_B}, $^6$Li~\cite{6Li_Selim_A,6Li_Selim_B,6Li_PenState_A,6Li_PenState_B}, and
$^{85}$Rb \cite{85Rb_JILA} are also consistent with values
of $a^{-}_{\rm 3b} /r_{\rm vdW}$ found for $^{133}$Cs ~\cite{133Cs_IBK_2}.
(Note that the work in Ref.~\cite{7Li_Kayk_B} also provided early suggestive
evidence of such universal behavior.) These observations provide strong evidence that the
three-body parameter has universal character for spherically-symmetric neutral atoms,
and therefore suggest that {\em something else} beyond the
universal theory needs to be understood.  

\begin{figure*}[htbp]
\includegraphics[width=\textwidth,angle=0,clip=true]{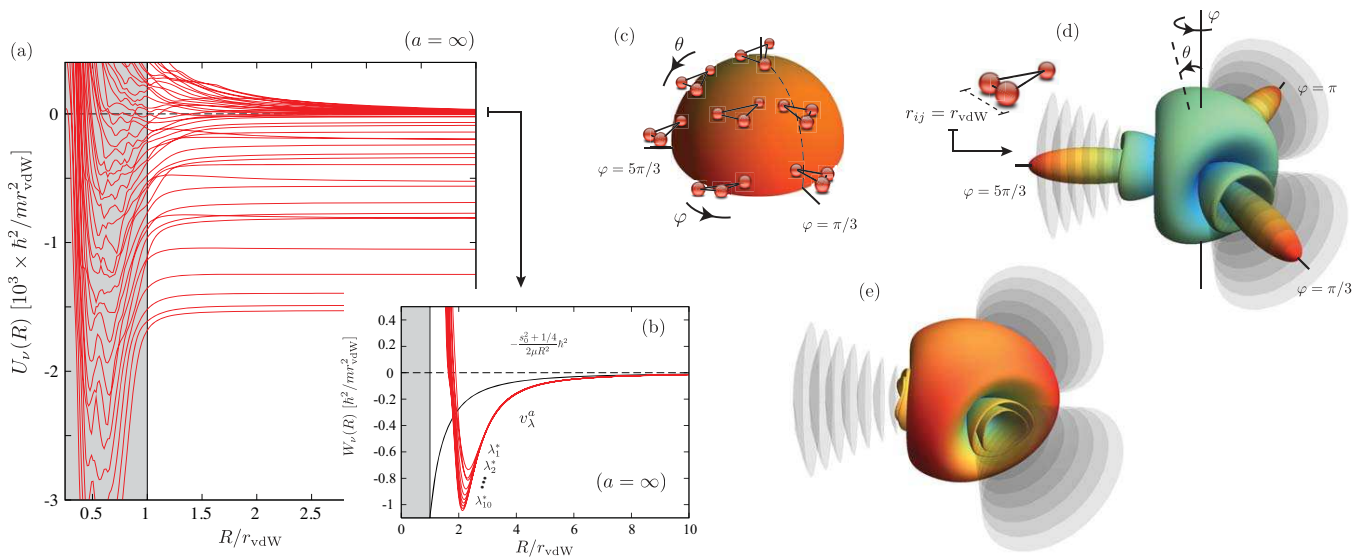}
\caption{
(a) The full energy landscape for the three-body potentials
at $a=\infty$ for our $v^{a}_{\lambda}$ model potential
supporting 25 two-body
bound states. 
(b) effective diabatic potentials $W_{\nu}$ relevant
for Efimov physics for $v_{\lambda}^a$ with an increasingly
large number of bound states ($\lambda^*_n$ is the value of $\lambda$
that produces $a=\infty$ and $n$ $s$-wave bound states). 
$W_{\nu}$ converge to a universal potential displaying the repulsive barrier 
at $R\approx2r_{\rm vdW}$ that prevents particles 
access to short distances.
(c)-(e) demonstrate the suppression of the wave function inside the
potential well 
through the channel functions $\Phi_{\nu}(R;\theta,\varphi)$  
for $R$ fixed near the minima of the Efimov potentials in (b).
(c) shows the mapping of the geometrical configurations onto the
hyperangles $\theta$ and $\varphi$. (d) and (e) show the 
channel functions, where the 
``distance'' from the origin determines $|\Phi_{\nu}|^{1/2}$, for two distinct cases: 
in (d) when there is a
substantial probability to find two particles inside the potential
well (defined by the region containing the gray disks)
and in (e) with a reduced probability 
 --- see also our discussion in Fig.~\ref{Fig2}.
In (d) and (e), we used the potentials $v_{\rm sch}$ 
and $v_{\lambda}^a$,
respectively, both with $n=3$.
Note that we have used $\Phi_\nu(R;\theta,\varphi)=\Phi_\nu(R;\pi-\theta,\varphi)$
to generate the lower half of the surface plots.
}
\label{Fig1}
\end{figure*}

In this Letter, we identify precisely the physics beyond the universal theory
needed to explain a universal three-body parameter, presenting theoretical
evidence to support the recent experimental observations.
Previous work has shown that the three-body
parameter can be universal --- that is, independent of the details of
the interactions --- in three-polar-molecule systems~\cite{Dipoles} and
in three-atom systems near narrow
Fano-Feshbach resonance~\cite{PetrovNarrowRes,Stoof}, although recent
work has shown that the latter case likely requires even more
finely-tuned conditions~\cite{OurNarrowRes}. Our present numerical
analysis, however, adds another, broader class of systems with a
universal three-body parameter: systems with two-body interactions that
efficiently suppress the probability to find any pair of particles
separated by less than $r_{0}$ 
(see Section~A in Ref.~\cite{EPAPS}).  
This class of systems, therefore, is more closely related to systems near broad
Fano-Feshbach resonances \cite{ChinReview}.

Such a suppression could derive from the usual
classical suppression of the probability for two particles to exist
between $r$ and $r+dr$ in regions of high {\em local} velocity $\hbar k_{L}(r)$, which
is proportional to $[\hbar k_{L}(r)/m dr]^{-1}$
($m$ being the particle mass), the time spent in that interval $dr$
(see Section B in Ref.~\cite{EPAPS}).  
It is possible that there could be an 
additional suppression as well,
through quantum reflection from 
a potential {\em cliff} \cite{QRefReview}. Systems supporting many bound
states, such as the neutral atoms used in ultracold experiments with
their strong van der Waals attraction, clearly exhibit this
suppression. In general, a finite-range two-body potential that
supports many bound states decreases steeply with decreasing
interparticle distance $r$, starting when $r/r_{\rm vdW}\lesssim1$,
at which point the potential cliff plays
a role analogous to a repulsive potential for low-energy
scattering. We demonstrate this fact by showing that the three-body 
parameter in the presence of many two-body bound states roughly 
coincides with that for a 100\% reflective two-body model potential, 
where the two-body short-range potential well is replaced by a hard-sphere. 

\begin{figure}[htbp]
\includegraphics[width=0.99\columnwidth,angle=0,clip=true]{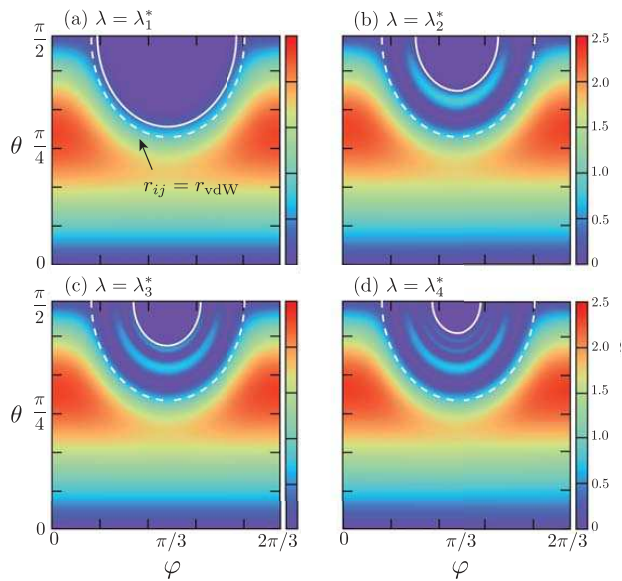}
\caption{
Density plot 
of the three-body probability density $|\Phi_{\nu}(R;\theta,\varphi)|^2 \sin2\theta$ which 
determines the three particle configuration [see Fig.~\ref{Fig1}~(c)] in the $\theta$-$\varphi$
hyperangular plane for a fixed $R$ ($\sin2\theta$ is the volume element).
(a)--(d) show the results for an $R$ near the
minimum of the Efimov potentials in Fig.~\ref{Fig1}~(b) for the
first four poles of the $v_\lambda^a$ model
as indicated. (a) shows that there is a negligible probability to find the
particles at distances smaller than $r_{\rm vdW}$ (outer dashed
circle) and, of course, inside the $1/r^{12}$ repulsive barrier
(inner solid circle). For higher poles, i.e., as the strength
of the hard-core part of $v_{\lambda}^a$ potential decreases, 
the potential becomes deeper and penetration into the region $r<r_{\rm
vdW}$ is now classically allowed. Nevertheless, 
(b)--(e) show that
inside-the-well suppression still efficiently prevents the particles 
from being found at distances $r<r_{\rm vdW}$.
In fact, we calculated the probability to find the atoms at $r<r_{\rm vdW}$ 
and found to be in the range $2\%$--$4\%$.
}
\label{Fig2}
\end{figure}

\begin{figure}[htbp]
\includegraphics[width=0.9\columnwidth,angle=0,clip=true]{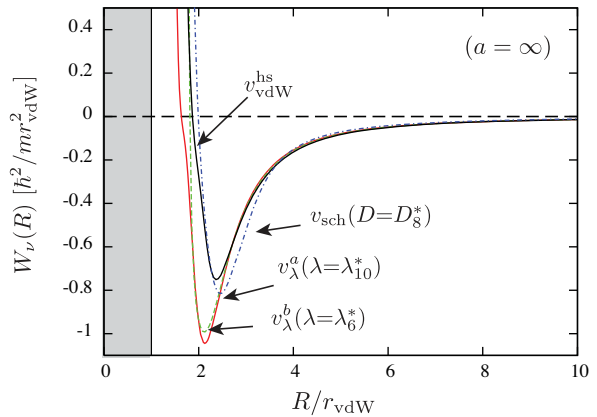}
\caption{
The Efimov potential obtained from the different two-body potential models used here.
The reasonably good agreement between the results obtained using models
supporting many bound states ($v_{\rm sch}$, $v^a_{\lambda}$ and $v_{\lambda}^b$) 
and $v_{\rm vdW}^{\rm hs}$ [obtained by replacing the deep potential
well with a hard wall but having only {\em one} 
(zero-energy) bound state] supports our conclusion that the
inside-the-well suppression of the wave function is the 
main physical mechanism behind the universality of the three-body
effective potentials. The differences between these potentials 
are shown to lead to differences of a few percent in the three-body parameter.
Note that the two-body potentials $v_{\lambda}^a$, $v_{\lambda}^b$, 
and $v_{\rm vdW}^{\rm hs}$ all have an asymptotic van der Waals tail, whereas $v_{\rm sch}$
decays exponentially at large distances.
}
\label{Fig3}
\end{figure}

\begin{figure*}[htbp]
\includegraphics[width=0.9\textwidth,angle=0,clip=true]{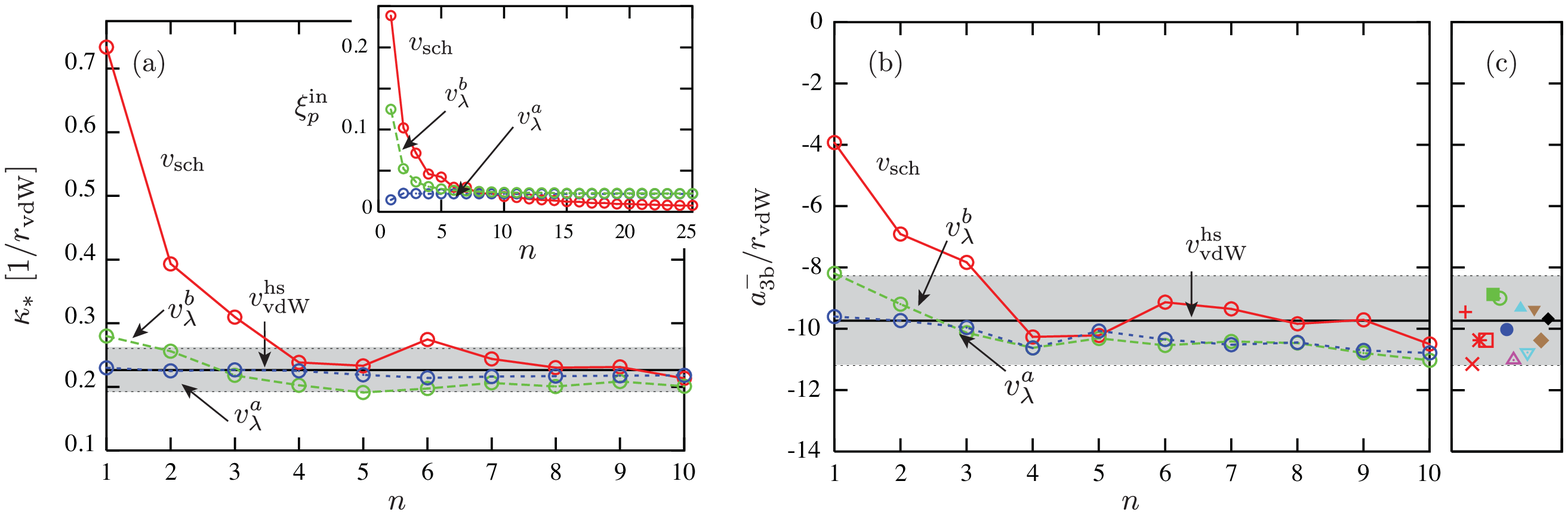}
\caption{
Values for the three-body parameter (a) $\kappa_*$ and (b) $a_{\rm 3b}^-$ as 
functions of the number $n$ of 
two-body $s$-wave bound states for each of the potential model studied here. 
The values for $a^-_{\rm 3b}$ were 
obtained by determining the value of $a$ at which the Efimov state becomes
unbound. (c) Experimental values for $a_{\rm 3b}^-$ for $^{133}$Cs \cite{133Cs_IBK_2} 
(red: $\times$, $+$, $\Box$, and $\ast$), $^{39}$K \cite{39K_LENS} (magenta: $\vartriangle$), 
$^{7}$Li \cite{7Li_Rice} (blue: $\bullet$) and \cite{7Li_Kayk_A,7Li_Kayk_B}
(green: $\blacksquare$ and $\circ$), $^6$Li \cite{6Li_Selim_A,6Li_Selim_B}
(cyan: $\blacktriangle$ and $\triangledown$) and \cite{6Li_PenState_A,6Li_PenState_B} 
(brown: $\blacktriangledown$ and $\lozenge$), and $^{85}$Rb \cite{85Rb_JILA} (black: $\blacklozenge$). 
The gray region specifies a band where there is a $\pm15\%$ 
deviation from the $v^{\rm hs}_{\rm vdW}$ results. 
The inset of (a) shows the suppression parameter $\xi^{\rm in}_{p}$ 
 [Eq.~(S.5) in Ref.~\cite{EPAPS}]  
which can be roughly understood as the degree of sensitivity to nonuniversal 
corrections. Since $\xi^{\rm in}_p$ is always finite --- even in the large $n$ limit --- nonuniversal effects 
associated with the details of the short-range interactions can still play an important role. 
One example is the large deviation in $\kappa_*$ found for the $v_{\rm sch}$ ($n=6$) model, caused by
a weakly bound $g$-wave state. 
For $n>10$ we expect 
$\kappa_*$ and
$a_{\rm 3b}^-$ to lie within the range of $\pm$15\% stablished for $n\le10$.
}
\label{Fig4}
\end{figure*}

The starting point for our investigation of the universality of the three-body
parameter is the adiabatic hyperspherical representation
\cite{DIncaoJPB,SunoHyper}. This representation offers a simple and
conceptually clear description by reducing the problem to the solution
of the ``hyperradial'' Schr{\"o}dinger equation :
\begin{multline}\left[-\frac{\hbar^2}{2\mu}\frac{d^2}{dR^2}+W_{\nu}(R)\right]F_{\nu}(R)
  \\+\sum_{\nu'\neq\nu} W_{\nu\nu'}(R) F_{\nu'}(R)=E
  F_\nu(R).\label{radeq}\end{multline}
Here, the hyperradius $R$ describes the overall size of the system; $\nu$
is the channel index; $\mu=m/\sqrt{3}$ is the three-body reduced mass
for particle masses $m$; $E$ is the total energy; and $F_{\nu}$ is the hyperradial wave
function. The nonadiabatic couplings $W_{\nu\nu'}$ drive inelastic
transitions, and the effective hyperradial potentials $W_{\nu}$ support
bound and  resonant states.  
To treat problems with deep two-body interactions 
--- necessary to see strong inside-the-well suppression --- requires us to
solve Eq.~(\ref{radeq}) for two-body model interactions 
that support many bound states, a challenge for most theoretical
approaches. Using our recently developed methodology
\cite{JiaSVD}, however, we were able to treat systems with up to 100
two-body ro-vibrational bound states
and solve Eq.~(\ref{radeq}) beyond the adiabatic approximation.  
Here, we explore the universality of the three-body parameter by
analyzing a number of model potentials [see $v_{\rm sch}$, $v^a_\lambda$, 
$v^b_\lambda$, $v_{\rm vdW}^{\rm hs}$ in Eqs.~(S.1)-(S.4) of Ref.~\cite{EPAPS}]
with different numbers of bound states. 

Figure \ref{Fig1}~(a) shows the adiabatic potentials 
$U_{\nu}$ at $|a|=\infty$ obtained using only the two-body Lennard-Jones potential, $v^a_\lambda$,
supporting 25 dimer bound states.
At first glance,
it is difficult to identify any universal properties of these
potentials. Efimov physics, however, occurs at a very small energy
scale near the three-body breakup threshold. Indeed, a closer analysis
of the energy range $|E|<\hbar^2/mr_{\rm vdW}^2$ [Fig.~\ref{Fig1}~(b)] reveals the
universal properties of the key potential curve controlling Efimov
states and universal scattering properties. 

Figure~\ref{Fig1}(b) shows one of our most important pieces of theoretical evidence for
universality of the three-body parameter: the effective adiabatic
potentials $W_{\nu}$ obtained using $v_{\lambda}^a$ for more and more two-body 
bound states converge to a single universal curve. 
[In some cases in Fig.~\ref{Fig1}~(b) we have manually
diabatized $W_{\nu}$ near sharp avoided crossings 
in order to improve the visualization 
 (see Section D in Ref.~\cite{EPAPS}).]  
As one would expect, the usual Efimov behavior for the effective potentials, 
$W_\nu$=$-\hbar^2(s_{0}^2+1/4)/2\mu R^2$ with $s_{0}\approx1.00624$, 
is recovered for $R>10 r_{\rm vdW}$. It is remarkable, however, that $W_{\nu}$ also converge to a 
universal potential for $R<10r_{\rm vdW}$ and, more importantly,  
these effective potentials display a repulsive wall or barrier at $R\approx2r_{\rm vdW}$. This 
barrier prevents the close collisions that would probe the short-range three-body physics,
including three-body forces known to be important in chemistry,
thus making the three-body parameter universal as we will confirm
below. This is in fact our most striking result: {\it
a sharp cliff or attraction in the two-body interactions produces a
strongly repulsive universal barrier in the effective three-body
interaction potential.}

{\em Qualitatively}, this universality derives from the reduced probability
to find particles inside the attractive two-body potential well. 
This effect can be seen in terms of the channel functions $\Phi_{\nu}$ 
  (see Section C in Ref.~\cite{EPAPS}), 
in Figs.~\ref{Fig1}~(c)-(e) and the hyperangular probability densities in Fig.~\ref{Fig2}.
In the adiabatic hyperspherical representation, the space forbidden 
to the particles fills an increasingly larger portion of the hyperangular 
volume as $R$ decreases.
This evolution can be visualized as the dashed lines in 
Fig.~\ref{Fig2} (a)--(d) expanding outward.  In the process, 
the channel function $\Phi_\nu$ is squeezed into an increasingly smaller volume, 
driving its kinetic energy higher and producing the repulsive barrier in the universal Efimov 
potential. Moreover, this suppression implies that the details of the
interaction should be largely unimportant. Consequently, different two-body model potentials 
should give  similar three-body potentials. Figure~\ref{Fig3} demonstrates this
universality by comparing $W_{\nu}$ obtained from 
different potential models supporting many bound states.
Perhaps more importantly, it compares them with the results obtained from 
the
two-body model $v_{\rm vdW}^{\rm hs}$ that replaces the deep well by a hard wall,
essentially eliminating
the probability of observing any pair of atoms at short distances. 
%
{\em Quantitatively}, however, the fact that the barrier occurs
only at $R\approx2r_{\rm vdW}$ indicates that universality might not
be as robust as in the cases discussed in
Refs.~\cite{Dipoles,PetrovNarrowRes,Stoof,OurNarrowRes}. It is thus
important to quantify the value of the three-body parameter 
to assess the size of nonuniversal effects. 

In principle, the three-body parameter could be defined in terms
of {\em any} observable related to the Efimov physics~\cite{BraatenReview}.
Two of its possible definitions are~\cite{BraatenReview}: 
the value of $a=a^{-}_{\rm 3b}<0$ at which the first Efimov resonance
appears in three-body recombination 
(see for instance Ref.~\cite{FirstK3Calcs}) and 
$\kappa_*=(m|E^{0}_{\rm 3b}|/\hbar^2)^{1/2}$,  where $E^{0}_{\rm 3b}$ 
is the energy of the lowest Efimov state at $|a|\rightarrow\infty$.
Our numerical results for $\kappa_{*}$ and $a_{\rm 3b}^-$ are shown in Figs.~\ref{Fig4}(a) 
and (b), respectively, demonstrating their universality in the limit of
many bound states. In fact, the values for $\kappa_{*}$ and $a_{\rm 3b}^-$ in this limit
differ by no more than 15\% from the 
$v_{\rm vdW}^{\rm hs}$ results --- $\kappa_*=0.226(2)/r_{\rm vdW}$ and
$a_{\rm 3b}^-=-9.73(3) r_{\rm vdW}$ [solid black line in
Fig.~\ref{Fig4}(a) and (b)] --- indicating, once again, that the
universality of the three-body parameter is dependent upon the suppression 
of the probability density within the two-body potential wells.
We note that our results for single channel two-body models should be applied
for broad Feshbach resonances.
Given this picture, we attribute the non-monotonic behavior of 
$\kappa_*$ and $a_{\rm 3b}^-$ in Fig.~\ref{Fig4} to the small, but finite, 
probability to reach short distances, thus introducing nonuniversal effects 
related to the details of two- and three-body forces including interactions 
with an isolated perturbing channel.
Nevertheless, our results
for $a^-_{\rm 3b}$ are consistent with the experimentally measured
value for $^{133}$Cs \cite{133Cs_IBK_1,133Cs_IBK_2}, 
$^{39}$K \cite{39K_LENS}, $^{7}$Li \cite{7Li_Rice,7Li_Kayk_A,7Li_Kayk_B},
$^6$Li \cite{6Li_Selim_A,6Li_Selim_B,6Li_PenState_A,6Li_PenState_B,Comment2},
and $^{85}$Rb \cite{85Rb_JILA}, all of which lie within about 15\%
of the $v_{\rm vdW}^{\rm hs}$ result. 
Curiously, if one simply averages the experimental values, then the
result differs from the $v_{\rm vdW}^{\rm hs}$ result by 
less than 3\%. 

Previous treatments have failed to predict the universality of the
three-body parameter for various reasons.
In treatments using zero-range interactions, for instance,
the three-body parameter enters as a free parameter to cure
the Thomas collapse~\cite{Thomas}, preventing any statement
about its universality.
Finite range models, like those used in some of our own treatments \cite{DIncaoJPB} 
[corresponding to the results for $v_{\rm sch}$ with 
$n=2$ and $3$ in Figs.~\ref{Fig4}~(a) and (b)],
have failed for lack of substantial suppression of the probability
density in the two-body wells.  This scenario, however,
should reflect better the situation for light nuclei having
few bound states and shallow attraction.
In contrast to Ref.~\cite{DIncaoJPB}, other models
\cite{UedaLi,UedaHe,Ueda,Stoof,Kohler,Lasinio,Schmidt} have found better
agreement with experiments. Our analysis of these treatments, 
however, indicates that the two-body models used have many
of the characteristics of our $v_{\rm vdW}^{\rm hs}$, therefore satisfying the
prerequisite for a universal three-body parameter. 
A recent attempt~\cite{Chin3BP} to explain the universality of the three-body
parameter avoided explicit two-body models altogether, using instead an
{\em ad hoc} hyperradial potential that bore little resemblance to our
numerical potentials in Fig.~\ref{Fig1}.  This {\em ad hoc} three-body potential displayed
strong attraction at short distances in contrast to our key finding, that
a cliff of attraction for two bodies
produces a universal {\em repulsive} barrier in the three-body system.
Consequently, even though a universal three-body parameter
was found in Ref.~\cite{Chin3BP}, the fundamental understanding provided by the approach is uncertain
 (see our discussion in Section D of Ref.~\cite{EPAPS}).

In summary, our theoretical examination shows that the
three-body parameter controlling many of the universal properties of
Efimov physics can be a universal parameter under certain circumstances
that should be realized in most ultracold neutral atom
experiments. Provided the underlying two-body short-range interaction supports 
a large number of bound states, or it has some other property leading to the 
suppression of the wave function at short distances, three-body properties associated with 
Efimov physics can be expected to be universal. 
While these arguments suggest universality also for the three-body parameter in heteronuclear 
systems that exhibit Efimov physics with only resonant interspecies two-body interactions, verifying 
this prediction is a high priority for future theory and experiment.
Equally important is the exploration of the relationship between 
$a<0$ and $a>0$ Efimov features --- currently a subject of a number of 
controversies \cite{133Cs_IBK_2} --- under the new perpective our present
work offers. 


\begin{acknowledgments}
The authors acknowledge stimulating discussions with Y. Wang,  
S. Jochim, M. Weidem\"uller, P. S. Julienne, and J. M. Hutson. 
We appreciate C. Chin communicating his ideas prior to publication and for
useful comments on our manuscript.
This work was supported by the U.S. National 
Science Foundation and by an AFOSR-MURI grant.

\end{acknowledgments}



\setcounter{equation}{0}
\setcounter{figure}{0}

\renewcommand{\theequation}{S.\arabic{equation}}
\renewcommand{\thefigure}{S.\arabic{figure}}

\section*{SUPPLEMENTARY MATERIAL}\label{SupInfo}

\subsection{Two-body model interactions}\label{Twobody}

Our theoretical model for the two-body interactions mimics the tunability of the
interatomic interactions via Fano-Feshbach resonances \cite{S_ChinReview}
by directly altering the strength of the interparticle interactions 
and, consequently, leading to the desired changes in $a$. In this work, we have considered
various model interactions in order to test the universality of our three-body calculations
--- we consider a result is universal if it is independent of the particular choice of the underlying 
two-body interaction. The models we have used for the two-body interactions are:
\begin{align}
&v_{\rm sch}(r)=-D{\rm sech}^2\left({r}/{r_{0}}\right),\label{PotSech}\\  
&v_{\lambda}^{a}(r)=-\frac{C_{6}}{r^6}\left(1-{\lambda^6}/{r^6}\right)\label{PotC6a},\\ 
&v_{\lambda}^{b}(r)=-\frac{C_{6}}{r^6}\exp\left(-\lambda^6/r^6\right)\label{PotC6b},\\ 
&v_{\rm vdW}^{\rm hs}(r) = B_{\rm hs}\Theta\left(r_{\rm hs}-r\right)-\frac{C_{6}}{r^{6}}\Theta\left(r-r_{\rm hs}\right).
\label{PotHSC6}
\end{align}
The potential model in Eq.~(\ref{PotSech}) is the modified P\"oschl-Teller potential,
where $D$ determines the potential depth; Eq.~(\ref{PotC6a}) is the usual Lennard-Jones potential; and
Eq.~(\ref{PotC6b}) is a dispersion potential with a soft wall at short range. 
In Eq.~(\ref{PotHSC6}), $\Theta(x)$ the step-function [$\Theta(x)=0$ for $x<0$
and $1$ elsewhere. In practice, however, we have used a smooth representation of the step-function
$\Theta$ in order to simplify our numerical calculations.] The potential in Eq.~(\ref{PotHSC6}), therefore, 
consists of a hard-sphere potential for $r<r_{\rm hs}$ ($B_{\rm hs}\gg C_{6}/r_{\rm hs}^6$)
and a long-range dispersion $-1/r^6$ potential for $r>r_{\rm hs}$.
In the present study, the parameters $D$ and $\lambda$ in Eqs.~(\ref{PotSech})-(\ref{PotC6b}) 
are adjusted to give the desired $a$ and number of bound states. For convenience, we denote the values of 
$D$ and $\lambda$ at which there exist zero-energy bound states ($|a|\rightarrow\infty$) as $D^*_n$ and 
$\lambda^*_n$, where $n$ corresponds to the number of $s$-wave bound  states. For the potential 
model in Eq.~(\ref{PotHSC6}), however, we adjusted $r_{\rm hs}$ to produce the changes in $a$, but we 
only performed three-body calculations near the first pole at $r_{\rm hs}\approx0.8828r_{\rm vdW}$.

The potential models in Eqs.~(\ref{PotC6a})-(\ref{PotHSC6}) have in common the same large $r$ 
behavior given by the van der Waals interaction, $-C_{6}/r^6$, with $C_6$ the van der Waals dispersion 
coefficient. For these potentials, the important length scale is the van der Waals length 
defined as $r_{\rm vdW}\equiv (2\mu_{\rm 2b} C_6/\hbar^2)^{1/4}/2$ 
\cite{S_Flambaum}, where $\mu_{\rm 2b}$ is two-body reduced mass. 
Therefore, in order to compare the results from these models to those for 
$v_{\rm sch}$, we define an equivalent $r_{\rm vdW}$ for $v_{\rm sch}$ through the relationship
between $r_{\rm vdW}$ and the effective range $r_{\rm eff}$ for $v_{\rm sch}$, namely,
$r_{\rm vdW}\approx r_{\rm eff}/2.78947$ \cite{S_Flambaum}, valid for values $|a|\gg r_{0}$. 
In fact, for $v_{\rm sch}$ we have found that $r_{\rm eff} (|a|=\infty)$ depends on the potential depth 
$D_{n}^*$, while for van der Waals type of interactions, such as those in Eqs.~(\ref{PotC6a})-(\ref{PotHSC6}), $r_{\rm eff}$ 
is fixed by $r_{\rm eff}\approx2.78947 (2\mu_{\rm 2b} C_6/\hbar^2)^{1/4}/2$.
Nevertheless, after rescaling the three-body results obtained from $v_{\rm sch}$ 
in terms of $r_{\rm vdW}$, we found good agreement with those obtained using $v_{\lambda}^a$, $v_{\lambda}^b$, 
and $v_{\rm vdW}^{\rm hs}$, thus indicating that the effective range is in fact the most relevant length 
scale in the problem. We chose, however, 
to keep our results in terms of $r_{\rm vdW}$ to make a closer analogy to the experiments
using alkali atoms. 

\subsection{Suppression of inside-the-well probability}\label{TwobodyS}

Our claim here is that the origin of the universality of the three-body parameter is related to
the suppression of the probability to find two particles at distances $r<r_{\rm vdW}$. This suppression leads
to the formation of the universal three-body potential barrier near $R\approx2r_{\rm vdW}$, thus
strongly suppressing the three-body wave function at 
shorter distances where the details of the two- and three-body interactions 
  are important [see Fig.~1~(b) of our main text].  
Next we explore the origin of this suppression at the two-body level. 

To gain some insight into the likelihood of finding particles inside the potential well, we start by
defining the following quantities, 
\begin{eqnarray}
\xi_{p}^{\rm in}(k)&=&\frac{1}{r_{0}}{\int_{0}^{r_{0}}|\psi_k(r)|^2{\rm d}r}, \label{XipIn}\\
\xi_{p}^{\rm out}(k)&=&\lim_{r\rightarrow\infty}\frac{1}{r-r_{0}}{\int_{r_0}^{r}|\psi_k(r)|^2{\rm d}r},\label{XilOut}
\end{eqnarray}
where $\psi_k(r)$ is the two-body scattering wave function at energy $E_{\rm 2b}=k^2/2\mu_{\rm 2b}$, defined such that
\begin{eqnarray}
\psi_{k}(r)\stackrel{r\rightarrow\infty}{=}\frac{\sin(kr+\delta)}{\sin\delta},\label{asympsol}
\end{eqnarray}
with $\delta\equiv\delta(k)$  the $s$-wave scattering phase shift.
This definition for $\psi_k(r)$, therefore, leads to a zero-energy ($k\rightarrow0$) wave function of 
the form: $\psi_0(r)=1-r/a$. [Note that in the above equations $r_{0}$ is the characteristic range of the two-body 
interaction. For the potential model in Eq.~(\ref{PotSech}), $r_{0}$ is just the quantity in the argument of the sech function,
while for the potential models in Eqs.~(\ref{PotC6a})-(\ref{PotHSC6}) it is defined to be $r_{0}=r_{\rm vdW}$].

\begin{figure}[htbp]
\includegraphics[width=\columnwidth,angle=0,clip=true]{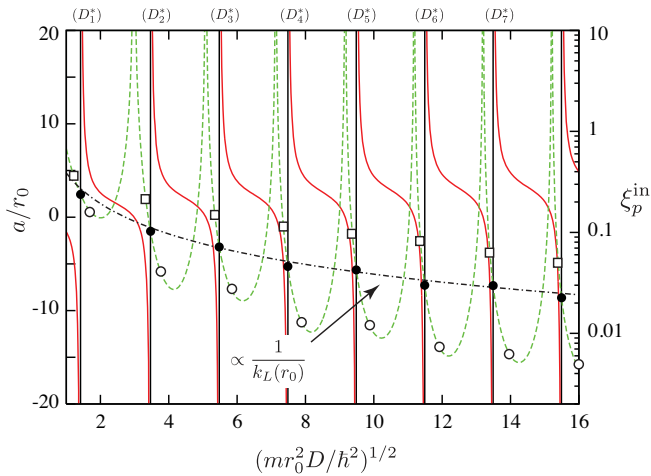}
\caption{
The red solid line
represents the scattering length, $a/r_{0}$, while the green dashed lines represents
the parameter $\xi_{p}^{\rm in}(k\rightarrow0)$. Both quantities are given as a function
of the depth $D$ of the two-body interaction model $v_{\rm sch}$ [Eq.~(\ref{PotSech})],
whose values for which $|a|=\infty$ are indicated in the figure as $D_{n}^*$, where $n$
is the number of $s$-wave states. Black circles, open circles and open squares are the values 
of $\xi_{p}^{\rm in}$ at $|a|\rightarrow\infty$, $a=5r_{0}$, and $a=-5r_{0}$, showing the suppression
of the $\xi_{p}^{\rm in}$ as the number of bound states increases.
The results for $\xi_{p}^{\rm in}$ also show higher efficiency of the inside-the-well suppression for 
$|a|/r_{0}\gg1$. The black dash-dotted line shows the semi-classical suppression factor $1/k_{L}(r_{0})$
demonstrating the origin of the suppression mechanism.
}
\label{Fig1A}
\end{figure}

The parameters $\xi_{p}^{\rm in}$ and $\xi_{p}^{\rm out}$ can be associated with the ``average'' amplitude of 
the wave function inside and outside the potential well, respectively. We, therefore, define the amplitude inside 
the well {\em relative to the amplitude outside the well} as:
\begin{eqnarray}
\xi_{p}^{\rm rel}(k)&=&\frac{\xi_{p}^{\rm in}(k)}{\xi_{p}^{\rm out}(k)}
=2\:\xi_{p}^{\rm in}(k)
\sin^2\delta.
\end{eqnarray}
The relative amplitude above vanishes in the limit $k\rightarrow0$ ($\sin\delta\approx -ka)$,
as a result of our choice for the asymptotic solution in Eq.~(\ref{asympsol}), 
except at $|a|=\infty$, when $\delta$ is an odd multiple of $\pi/2$.
The quantity $\xi_{p}^{\rm in}(k)$, however, remains finite in the $k\rightarrow0$ limit.
We believe, therefore, that  $\xi_{p}^{\rm in}(k\rightarrow0)$ to be the most relevant parameter for our analysis 
on the inside-the-well suppression. 
Rigorously speaking, $\xi_{p}^{\rm in}$ is {\em not} a probability, but it does measure the likelihood 
of finding two particles within $r<r_{0}$, where the short-range interactions are experienced. 
Figure \ref{Fig1A} shows a typical result for $|a|/r_{0}$ and $\xi_{p}^{\rm in}(k\rightarrow0)$ for the 
two-body potential in Eq.~(\ref{PotSech}).
Figure~\ref{Fig1A} shows that in the universal regime near the poles in $a$,
the wavefunction is suppressed (small $\xi_{p}^{\rm in}$) and that this suppression
becomes more efficient as the potential becomes deeper and more states are bound. 
The black filled circles, open circles and open squares in Fig.~\ref{Fig1A}, showing the 
values of $\xi_{p}^{\rm in}$ at $|a|\rightarrow\infty$, $a=5r_{0}$, and $a=-5r_{0}$, respectively, 
illustrate this trend. Note, however,
that for values $|a|\lesssim r_{0}$, the parameter $\xi_{p}^{\rm in}$ quickly increases, indicating a higher likelihood to 
find particles inside the potential well. Similar results are also obtained for the potentials $v_{\lambda}^a$ and $v_{\lambda}^b$ 
[Eqs.~(\ref{PotC6a}) and (\ref{PotC6b}), respectively]. 

The suppression in $\xi_{p}^{\rm in}$ (for a fixed value of $|a|\gg r_{0}$) can be understood to be a result
of the usual semiclassical suppression of the wave function. In the WKB approach 
\cite{S_BerryReview,S_Gribakin,ChrisRMP}, the wave function inside the potential well can be written as,
\begin{equation}
\psi_{\rm WKB}(r<r_{0})=\frac{C}{k_{L}^{1/2}(r)}\sin\left[\int^{r}k_{L}(r')dr'+\frac{\pi}{4}\right],\label{psiWKB}
\end{equation}
where $C$ is a normalization constant and $k_{L}^2(r)=2\mu_{\rm 2b}[E_{\rm 2b}-v(r)]$ defines the {\em local}
wave number $k_{L}(r)$. Therefore, for deep potentials, the suppression of the wave function inside the potential well
is simply related to factor $k_{L}(r)^{-1/2}$ which leads to the suppression of the amplitude of
the WKB wave function [Eq.~(\ref{psiWKB})] between $r$ and $r+dr$. Physically, this can be seen as the increase 
of the {\em local} velocity $\hbar k_{L}(r)/m$ ($m$ being the particle mass) and, therefore, the decrease of the time spent in that interval $dr$,
$[\hbar k_{L}(r)/mdr]^{-1}$. Therefore, in the WKB approximation, one would expect $\xi_{p}^{\rm in}$ 
to be simply proportional to $1/k_{L}$. We test this expectation by plotting in Fig.~\ref{Fig1A}
the value of $1/k_L(r_{0})$ (black dash-dotted line) which shows that the suppression in 
$\xi_{p}^{\rm in}(k\rightarrow0)$ is consistent with the semiclassical suppression described above.

We have also considered the possibility that two-body quantum reflection \cite{S_QRefReview,S_QRef} could be
responsible for the suppression of inside-the-well probability, and consequently the universality of 
the three-body parameter. Quantum reflection is a fundamental phenomenon involving reflection of a 
low energy quantum matter wave from a potential {\it cliff}. 
In the context of ultracold collisions of ground state neutral atoms, quantum reflection can occur when two atoms 
approach within a distance comparable to the range of the interaction, $r_{0}$, and it is caused by the rapid change 
of the potential in this region \cite{S_QRef}. In that case, just a fraction of the incoming wave probes the details of the 
atom-atom interaction at $r<r_{0}$. Nevertheless, our analysis leads us to believe that this is of secondary 
importance. The main reason is that in a three-body system,  even at zero (three-body) energy, two particles can have energies 
that are not low enough to ensure that quantum reflection plays a rule. In fact, accordingly to our estimates, in a three-body 
system the average two-body energy will be proportional to $1/2\mu R^2$. This implies that near the universal 
barrier ($R\approx2r_{\rm vdW}$) the two-body energy can be of the order of $1/mr_{\rm vdW}^2$. 
Based on a model similar to the one used in Ref.~\cite{S_QRef} the reflection probability at such 
energies is about 10\%, which is too low to explain our observations. 
Our above analysis, however, shows that for such energies $\xi_{p}^{\rm in}(k)$ does not differs substantially
from $\xi_{p}^{\rm in}(0)$, offering a more consistent picture than quantum reflection.
Evidently, to precisely quantify the importance of quantum reflection requires 
a much more detailed analysis of how the energy of two particles in a three-body system behaves as a function of $R$, 
which is beyond the scope of our present study.

\subsection{Three-body hyperspherical adiabatic representation}\label{Threebody}

As mentioned in our main text, our analysis of the universality of the three-body parameter
was based on the hyperspherical adiabatic representation. The following, is a sketch of its
basic aspects and fundamental equations. A more detailed description
can be found, for instance, in Refs.~\cite{S_DIncaoJPB,S_SunoHyper}. We start here from 
the hyperradial Schr{\"o}dinger equation :
\begin{multline}\left[-\frac{\hbar^2}{2\mu}\frac{d^2}{dR^2}+W_{\nu}(R)\right]F_{\nu}(R)
  \\+\sum_{\nu'\neq\nu} W_{\nu\nu'}(R) F_{\nu'}(R)=E
  F_\nu(R).\label{Sradeq}
\end{multline}
The hyperradius $R$ can be 
expressed in terms of interparticle distances for a system of three equal masses $m$,  
$R=3^{-1/4}(r_{12}^2+ r_{23}^2+r_{31}^2)^{1/2}$, and describes the overall size of the system; 
$\nu$ is a collective index that represents all quantum numbers necessary to
label each channel; $\mu=m/\sqrt{3}$ is the three-body reduced mass;
$E$ is the total energy; and $F_{\nu}$ is the hyperradial wave
function. The simple picture resulting from this representation originates
from the fact that nonadiabatic couplings $W_{\nu\nu'}$ are the quantities
that drive inelastic transitions, and that the effective hyperradial potentials 
$W_{\nu}$, like any usual potential, support bound and resonant states, but now for
the three particle system.  

To obtain $W_{\nu}$ and $W_{\nu\nu'}$, however,
one needs to diagonalize the adiabatic Hamiltonian $H_{\rm ad}$
which includes the hyperangular kinetic energy and {\em all} of the interactions
~\cite{S_DIncaoJPB,S_SunoHyper}. Our present study 
has assumed that the interactions are given as a pairwise sum of two-body model 
potentials and used those from Eqs.~(\ref{PotSech})-(\ref{PotHSC6}). 
The eigenvalues of $H_{\rm ad}$, determined for fixed
values of $R$ by using the method in Ref.~\cite{S_SunoHyper}, are the three-body potentials $U_{\nu}(R)$ 
  [main text Fig.~1~(a)]  
associated to the respective channel functions (eigenfunctions) $\Phi_{\nu}(R;\Omega)$ \cite{S_PhiRefs_A,S_PhiRefs_B}. 
Therefore, the channel functions $\Phi_{\nu}$ contain all the information necessary to determine,
for fixed values of $R$, the geometric structure of the three particle system.
For our present study concerning three identical bosons in the total angular momentum state 
$J^{\pi}=0^+$ ($\pi$ being the parity), the set of hyperangles
$\Omega$ \cite{S_SunoHyper} is reduced to simply $\{\theta,\varphi\}$. This helps to visualize the 
geometric structure of the three-body system 
  [see Figs.~1~(c)-(e) and Fig.~2 of our main text].

Finally, to determine the effective potentials $W_{\nu}$ and nonadiabatic couplings $W_{\nu\nu'}$,
which are the main quantities in Eq.~({\ref{Sradeq}}), one can use the expressions,
\begin{eqnarray}
W_{\nu}(R)&=&U_{\nu}(R)-\frac{\hbar^2}{2\mu}Q_{\nu\nu}(R),\label{EffPot}\\
W_{\nu\nu'}(R)&=&-\frac{\hbar^2}{2\mu}\left[2P_{\nu\nu'}(R)\frac{d}{dR}+Q_{\nu\nu'}(R)\right]\label{PQs},
\end{eqnarray}
where
\begin{eqnarray}
P_{\nu\nu'}(R)&=&\langle\!\langle \Phi_\nu(R;\Omega)| \frac{\partial}{\partial R}|
\Phi_{\nu'}(R;\Omega)\rangle\!\rangle,\label{Pc}\\
Q_{\nu\nu'}(R)&=&\langle\!\langle \Phi_\nu(R;\Omega)| \frac{\partial^2}{\partial R^2}|
\Phi_{\nu'}(R;\Omega)\rangle\!\rangle.\label{Qc}
\end{eqnarray}
In the above expressions, the double-bracket matrix elements indicate that integrations
are carried out over only the hyperangles $\Omega$. 
As is mentioned in the main text, to treat problems with deep two-body potentials,
and consequently many bound states, we have used a modified version of the above
formulation since $P_{\nu\nu'}$ and $Q_{\nu\nu'}$, defined above, can become 
numerically unstable. This problem can be overcome using the formulation 
developed in Ref. \cite{S_JiaSVD}. Apart from this more technical aspect, the above 
formulation gives the general ideas concerning the hyperspherical representation. 
We encourage the interested reader to look for more details in 
Refs.~\cite{S_DIncaoJPB,S_SunoHyper,S_JiaSVD}.  

\subsection{Effective adiabatic potentials} \label{ThreebodyPot}

\begin{figure}[htbp]
\includegraphics[width=\columnwidth,angle=0,clip=true]{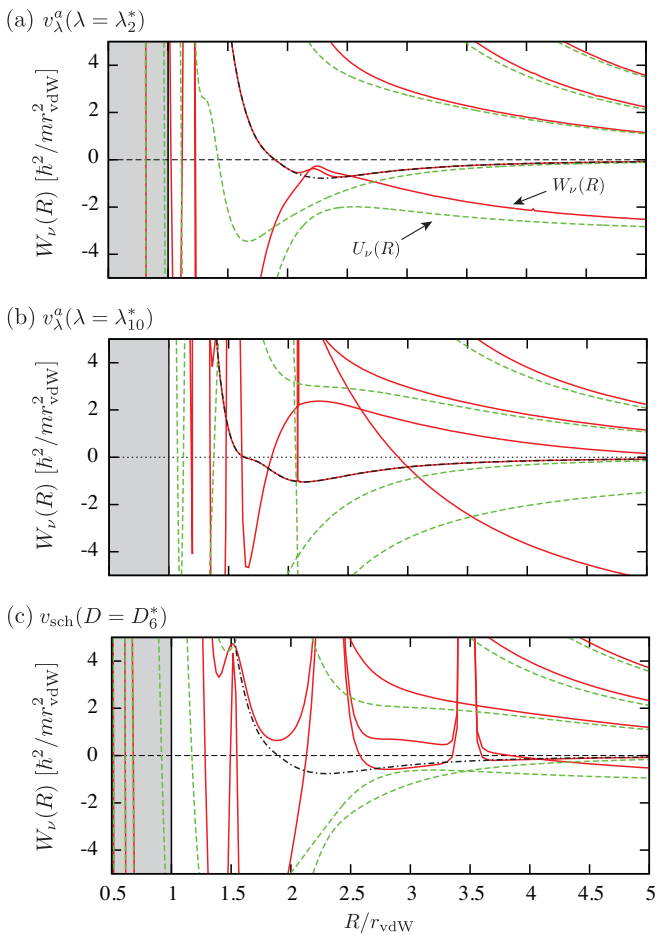}
\caption{
Comparison of $U_{\nu}(R)$ (green
dashed curves) and $W_{\nu}(R)$ (red solid curves) demonstrating the importance of nonadiabatic effects 
introduced by $Q_{\nu\nu}(R)$ in Eq.~(\ref{EffPot}). (a) and (b) show the results for the potential models 
$v_{\lambda}^a(\lambda=\lambda_2^*)$ and $v_{\lambda}^a(\lambda=\lambda_{10}^*)$ 
[Eq.~(\ref{PotC6a})], respectively, and in (c) we show the results obtained for $v_{\rm sch}(D=D_{6}^*)$ 
[Eq.~(\ref{PotSech})]. (a) and (b) also show the effect of the diabatization scheme used in order 
to prepare some of our figures in the main text (dash-dotted curves). The goal of the diabatization
is to eliminate the sharp features resulting from $Q_{\nu\nu}(R)$ in Eq.~(\ref{EffPot}) which 
should not contribute substantially to the three-body observables. The case 
shown in (c), however, does not allow us to easily trace the diabatic version of the potentials 
$W_{\nu}$. In this case, however, the ``less'' sharp features have a larger contribution
due to the crossing with a three-body channel describing a collision between a $g$-wave molecular
state and a free atom, 
giving rise to the anomalous $n=6$ point for the $v_{\rm sch}$ model 
  in Fig.~4 of the main text.
Although such cases are relatively infrequent in our calculations (and occur
mostly for the $v_{\rm sch}$ model), they do illustrate nonuniversal effects that can 
affect the three-body parameter. Nevertheless, the three-body observables obtained for cases like the one 
shown in (c) are still within the 15\% range we claimed for the universality of the three-body parameter.}
\label{Fig2A}
\end{figure}

 The effective potentials shown in Fig. 1~(a) of the main text
are very complicated, making identification of the important physics challenging.
For that reason, this section presents some details of our work that not only
give support to our physical interpretation of the nature of the three-body parameter 
but also show how some important nonuniversal aspects appear in the hyperspherical adiabatic 
representation. In fact, Fig.~\ref{Fig2A} shows some of the most drastic nonadiabatic effects 
found in our calculations. 
The model proposed in Ref.~\cite{S_Chin3BP} is also considered, and we show that
this model offers an interesting qualitative picture; but since
incorporation of corrections to that model diminishes its accuracy in
the three-body parameter universality, its agreement with experiment
might be fortuitous.

We first consider the validity of the single-channel adiabatic hyperspherical approximation
and point out the manner in which some important nonuniversal features manifest themselves.
Figure \ref{Fig2A} shows the results for $U_{\nu}(R)$ and $W_{\nu}(R)$ obtained from three different two-body 
potential models. Figures~\ref{Fig2A}~(a) and (b) show the results for the potential models 
$v_{\lambda}^a(\lambda=\lambda_2^*)$ and $v_{\lambda}^a(\lambda=\lambda_{10}^*)$ 
[Eq.~(\ref{PotC6a})], respectively, while Fig.~\ref{Fig2A}~(c) shows the results obtained 
for $v_{\rm sch}(D=D_{6}^*)$ [Eq.~(\ref{PotSech})]. 
The most striking aspect of these figures is that $U_{\nu}(R)$ and $W_{\nu}(R)$ are 
substantially different, meaning that the nonadiabatic couplings $P_{\nu\nu'}(R)$ and 
$Q_{\nu\nu'}(R)$ [Eqs.~(\ref{Pc}) and (\ref{Qc})] are important near $R=r_{\rm vdW}$. Therefore, 
it is clear that one needs to go beyond a single channel approximation
in order to describe the three-body observables. It is worth noting that the nonadiabatic couplings 
originate from the hyperradial kinetic energy. Their large values are thus consistent with 
our physical picture in which the {\em local} kinetic energy increases and generates the 
repulsive barrier in our effective potentials.
It is also worth mentioning that since the three-body repulsive barrier prevents particles from approaching 
to small distances, the question of whether or not the short range physics actually changes
as a function of the external magnetic field (as in experiments in ultracold quantum gases) \cite{S_ChinArXiv}
can not be directly answered by observing features related to Efimov physics.

\begin{figure}[htbp]
\includegraphics[width=0.9\columnwidth,angle=0,clip=true]{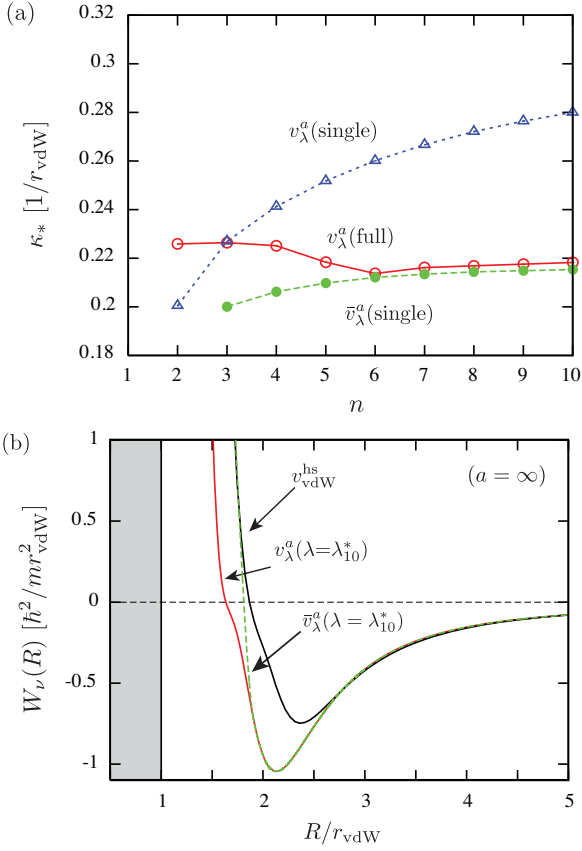}
\caption{
(a) This figure compares the energies (as characterized by the three-body parameter $\kappa_*$)
obtained from a single channel approximation 
with our full calculations. The three-body parameter $\kappa_{*}$ is shown for the 
$v_{\lambda}^a$ model in the single channel approximation (open triangles) 
as well as for our full numerical results (open circles). The single channel approximation
can be improved by imposing a simple change in the adiabatic potentials near 
the barrier, as is shown in (b). There we smoothly connect the potential for $v_{\lambda}^a$ (red
solid line) to the barrier obtained for $v_{\rm vdW}^{hs}$ (black solid line), resulting  
in the potential labeled by $\bar{v}_{\lambda}^a$ (green solid line). This new potential is actually more
repulsive and has energies [filled circles in (a)] that are much closer to our full numerical calculations. 
}
\label{Fig3A}
\end{figure}

The strong multichannel nature of the problem can be illustrated by comparing the 
results obtained from a single channel approximation to
Eq.~(\ref{Sradeq}), i.e., $W_{\nu\nu'}(R)=0$ ($\nu\ne\nu'$), with our solutions of the fully coupled system of equations.
Figure~\ref{Fig3A}~(a) shows the three-body parameter $\kappa_{*}$ [related to the energy of the lowest Efimov
state through the relation $\kappa_*=(m E/\hbar^2)^{1/2}$] obtained for the 
$v_{\lambda}^a$ model obtained in the single channel approximation (open triangles) 
as well as our full numerical results (open circles). The disagreement between 
these quantities increases with the number of $s$-wave bound states $n$, meaning that 
the physics controlling the results becomes more and more multichannel in nature. 
Nevertheless, we find that by imposing a simple change in the adiabatic potentials near 
the barrier --- to make the barrier {\em more} repulsive [see Fig.~\ref{Fig3A}~(b)] ---
the single channel approximation for $\kappa_{*}$ [filled circles in Fig.~\ref{Fig3A}~(a)] 
reproduces the full numerical calculations much better. This agreement indicates that most of the 
nonadiabaticity of the problem is related to the exact shape of the barrier and that, to some extent, 
the effect of the nonadiabatic couplings is to make the effective potential $W_{\nu}$ more repulsive.
For these reasons and, of course, the universality of our full calculations (see for instance 
  Fig.~4 of our main text), we believe that the short-range barrier in the three-body effective potentials  
indeed offers a physically valid explanation of the universality of the three-body parameter. 
  We emphasize, though, that Fig.~4 in the main text only includes the results from
our essentially exact solutions of the full calculations. The single channel results discussed here are
intended only as support of our physical interpretation.

\begin{figure}[htbp]
\includegraphics[width=\columnwidth,angle=0,clip=true]{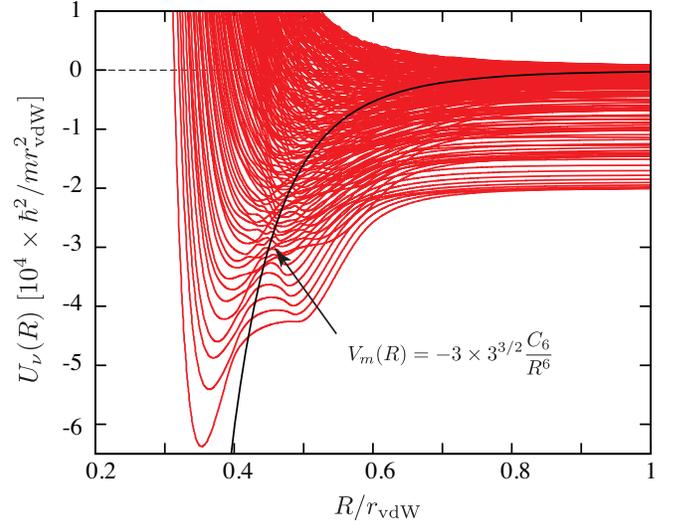}
\caption{
This figure shows the three-body 
potentials obtained using the $v_{\lambda}^{a}(\lambda=\lambda_{10}^*)$ model supporting a total of 
100 bound states. 
Roughly speaking, the potential of Eq.~(\ref{VmPot})~\cite{S_Chin3BP} (black solid line) can be seen as a diabatic potential 
since it passes near one of the series of avoided crossings.
}
\label{Fig4A}
\end{figure}

It is within this context that we analyzed the model proposed in Ref.~\cite{S_Chin3BP}.
In Ref.~\cite{S_Chin3BP}, the three-body effective potential important for Efimov physics was 
estimated by considering the different aspects controlling the physics at small and large distances.
At distances comparable to $R=r_{\rm vdW}$, it was assumed that the effective three-body potential is 
dominated by the contributions
from equilateral triangle geometries and included only on two-body interactions. Under these
assumptions, $r_{12}=r_{23}=r_{31}=r$ giving $R=3^{1/4}r$ (note that our 
definition for $R$ differs from that used in Ref.~\cite{S_Chin3BP}), 
the effective potential can be written as
\begin{eqnarray}
V_{m}(R)&=&-C_{6}/r_{12}^6 -C_{6}/r_{23}^6 -C_{6}/r_{31}^6\nonumber\\
             &=&-3C_{6}/r^6=-3\times3^{3/2}C_{6}/R^6\label{VmPot}.
\end{eqnarray}
This potential is expected to be valid for distances 
$R<\bar{A}=4\pi \Gamma(1/4)^{-2}3^{1/4}3^{3/8}r_{\rm vdW}\approx1.9r_{\rm vdW}$ 
\cite{S_Flambaum}.  With our method, however, we have the means to 
check the validity of Eq.~(\ref{VmPot}) by comparing it with our numerically calculated potentials.
Figure \ref{Fig4A} shows the three-body potentials obtained using the 
$v_{\lambda}^{a}(\lambda=\lambda_{10}^*)$ model supporting a total of 100 two-body bound states. 
The potential from Eq.~(\ref{VmPot}) is the black solid line passing near the series of 
avoided crossings
and might be loosely viewed as diabatically connecting 
the fully numerical potentials. This relation is reasonable given that this
sequence of avoided crossings has been shown in Ref.~\cite{S_Blume} 
to be related to the transition of the system from an equilateral triangle geometry
to other geometries. 
This figure therefore suggests that approximating the short range physics by Eq.~(\ref{VmPot}) is 
not wholly unphysical, 
but its validity depends on a strong assumption of diabaticity through a large number of avoided
crossings and
is thus probably {\em not} an approximation satisfactory for a more quantitative analysis.

The potential in Eq.~(\ref{VmPot}), however, was not actually used in the calculations in 
Ref.~\cite{S_Chin3BP}. Instead, it was used to motivate a claim that 
three-body quantum reflection plays an important role, 
allowing the short range behavior to be replaced by a repulsive potential for distances $R<\bar{A}\approx1.9r_{\rm vdW}$. 
It is interesting to note that the value $R\approx1.9r_{\rm vdW}$ obtained from Ref.~\cite{S_Chin3BP} for 
the position of the hard wall is quite close to the hyperradius where our potentials exhibit the 
universal barrier (see Fig.~\ref{Fig2A}, for instance), indicating that $\bar{A}$ might have some physical
meaning. It is worth mentioning, however, that the barrier we observed in our calculations is model independent,
i.e., it doesn't rely on the particular model used for the two-body interaction. 
For distances $R>\bar{A}$, the model in Ref.~\cite{S_Chin3BP} assumed the three-body effective potentials to be 
given by the universal Efimov formula, 
\begin{eqnarray}
V_{E}(R)&=&-\hbar^2\frac{s_{0}^2+1/4}{2\mu R^2}. \label{EfimovPot}
\end{eqnarray}
It is well known \cite{S_Efimov}, however, that this potential is valid for $r_{\rm vdW}\ll R \ll |a|$,
and $R=1.9r_{\rm vdW}$ is certainly out of this range. In fact, 
one can see in Fig.~1~(b) of the main text,  
that the use of the Efimov potential [Eq.~(\ref{EfimovPot})] for $R<10r_{\rm vdW}$ 
is a crude approximation. Nevertheless, using this model, Ref.~\cite{S_Chin3BP}
obtains $a_{\rm 3b}^-\approx-9.48r_{\rm vdW}$, a value consistent with experiments 
\cite{S_133Cs_IBK_1,S_133Cs_IBK_2,S_39K_LENS,S_7Li_Rice,S_7Li_Kayk_A,S_7Li_Kayk_B,S_6Li_Selim_A,S_6Li_Selim_B,
 S_6Li_PenState_A,S_6Li_PenState_B,S_85Rb_JILA} as well as with our present calculations for $a_{\rm 3b}^-$. 
However, extending the model of Ref.~\cite{S_Chin3BP} to the limit $|a|=\infty$, we obtained
$\kappa_{*}\approx0.037/r_{\rm vdW}$. Our result, by way of contrast, using 
$v_{\rm vdW}^{\rm hs}$ [Eq.~(\ref{PotHSC6})] is $\kappa_{*}=0.226(2)/r_{\rm vdW}$. 
We also have tested the effects of finite $a$ corrections on the model of Ref.~\cite{S_Chin3BP} by replacing Eq.~(\ref{EfimovPot})
with the three-body potential obtained with a zero-range model of the two-body interactions \cite{S_GreenF}. 
These corrections are particularly important near $R=|a|$. With this modification,
the model of Ref.~\cite{S_Chin3BP} leads to $a_{\rm 3b}^-\approx-39.96r_{\rm vdW}$.
For these reasons, we believe that this model's agreement with our results and experimental data is fortuitous.

\begin{figure}[htbp]
\includegraphics[width=0.9\columnwidth,angle=0,clip=true]{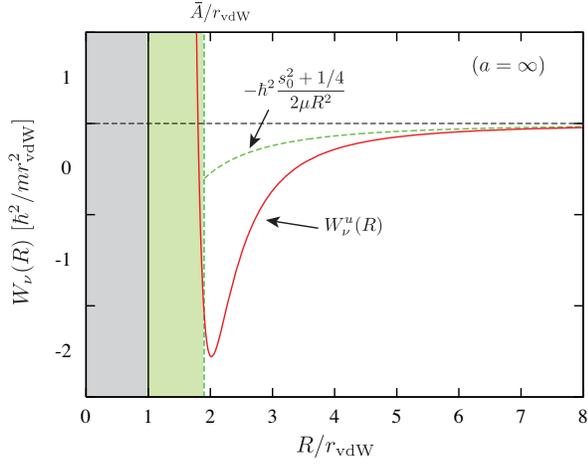}
\caption{
Comparison between the effective potential obtained from Ref.~\cite{S_Chin3BP}
and the potential from Eq.~(\ref{UnivPot}). 
}
\label{Fig5A}
\end{figure}

While multichannel couplings are needed to quantitatively describe this system, it is possible 
to construct an effective hyperradial potential curve that correctly describes the behavior of 
three atoms in the universal regime.  Such a potential curve could be useful for simplified future studies.  
The approximate form obtained from the present study is:
\begin{align}
\frac{2 \mu r_{\rm vdW}^2}{\hbar^2} &W^{u}_{\nu}(R) \approx
-\frac{s_{0}^2+1/4}{(R/r_{\rm vdW})^2}\nonumber\\
&-\frac{2.334}{(R/r_{\rm vdW})^3}
-\frac{1.348}{(R/r_{\rm vdW})^4}\nonumber\\
&-\frac{44.52}{(R/r_{\rm vdW})^5}+\frac{4.0\times10^4}{(R/r_{\rm vdW})^{16}}.\label{UnivPot}
\end{align}
Here $\mu$ is the three-body reduced mass and $r_{vdW}$ is the two-body van der Waals length.  
For comparison the speculative potential curve proposed by Chin \cite{S_Chin3BP} is shown, 
which does not resemble the present result even qualitatively at small distances, as it exhibits
far too little attraction in the region $R = 2 - 5 r_{\rm vdW}$.


\begin{thebibliography}{99}
\bibitem{CommentAmean}
The van der Waals length is defined as $r_{\rm vdW}\equiv (2
\mu_{\rm 2b} C_6/\hbar^2)^{1/4}/2$ where $C_{6}$ is the van der Waals
coefficient and the two-body reduced mass $\mu_{\rm 2b}$.
Note also that in Ref~\cite{133Cs_IBK_2} the results were quoted in terms of the mean scattering length 
$\bar{a}\approx0.9556 r_{\rm vdW}$ as defined in Ref. \cite{Flambaum}.
\end{thebibliography}
\end{document}